# Predicting the Future Performance of the Planned Seismic Network in Mainland China

**Jiawei Li[1], Arnaud Mignan[1,2], Didier Sornette[1,3], and Yu Feng[1]**

[1]Institute of Risk Analysis, Prediction and Management (Risks-X), Academy for Advanced Interdisciplinary Studies, Southern University of Science and Technology (SUSTech), Shenzhen, China.

[2]Department of Earth and Space Sciences, Southern University of Science and Technology (SUSTech), Shenzhen, China.

[3]Department of Management, Technology and Economics (D-MTEC), Swiss Federal Institute of Technology in Zürich (ETH Zürich), Zürich, Switzerland

Corresponding author: Jiawei Li (lijw@cea-igp.ac.cn)

**Key Points:**

- The spatial distribution of completeness magnitude for the new broadband seismic network in China is predicted.

- The completeness magnitude of China will soon decrease to 2.0, which means approximately 3 times more earthquakes available per year.

- The new seismic network will achieve the goal of 99% coverage for optimal earthquake prediction research based on seismic precursor.

**Abstract**

The new broadband seismic network in China will increase the number of stations from approximately 950 to 2000. The higher-resolution monitoring of the frequent smaller earthquakes expected inside Mainland China can be quantified via the completeness magnitude ($M_c$) metric. Using the Bayesian Magnitude of Completeness (BMC) method, we generate the spatial distribution of $M_c$ predicted for the new network, based on the prior model calibrated on the current earthquake catalog (2012 to 2021) and network configuration. If 99% of Mainland China is at present covered down to $M_c = 2.7$, this threshold will soon fall to $M_c = 2.0$. This means approximately 3 times more earthquakes available per year. Based on the observation that seismic precursors are most likely to be observed at least at 3 units below the mainshock magnitude, the new seismic network shall achieve the goal of almost total coverage for optimal seismic-based earthquake prediction research.

**Plain Language Summary**

The China Earthquake Administration (CEA) has currently launched an ambitious nationwide seismic network project, which will increase the number of stations from approximately 2,000 to 15,000 in total, from 950 to 2000 for the broadband seismic stations used to compile earthquake catalog. The new network is planned to go online by the end of 2022. In half of Mainland China, the inter-station distance will soon be smaller than 100 km, 23% be 50 km, and 3% be 25 km. Of all possible ways to characterize the higher-resolution monitoring of the frequent smaller earthquakes expected inside Mainland China, the completeness magnitude ($M_c$) remains one of the most commonly used. The completeness magnitude $M_c$ metric is defined as the smallest magnitude value above which the Gutenberg-Richter law is validated, or in other words, as the threshold above which earthquakes in a given space-time volume are detected with a probability tending to one. Using the prior model of the Bayesian Magnitude of Completeness (BMC) method calibrated on the Chinese earthquake catalog from January 1, 2012 to July 11, 2021, we predict the spatial distribution of $M_c$ for the new network based on the planned network configuration. If almost the entire Mainland China is at present covered down to $M_c = 2.7$, this threshold will fall to $M_c = 2.0$ in the near future. This means approximately 3 times more earthquakes will be recorded in the complete catalog available for statistical analysis per year (for $a = 6.76$ and $b = 0.85$ in the Gutenberg-Richter law $\log_{10} N = a - b \cdot M_c$). Based on the observation that abnormal seismicity as precursors are most likely to be observed at least at 3 units below the mainshock magnitude, and assuming earthquakes to be potentially damaging at $M \geq 5$, the new seismic network shall achieve the goal of 99% coverage for optimal seismic-based earthquake prediction research.

**1 Introduction**

The densification of a seismic network improves earthquake detectability to smaller magnitudes (Engdahl and Bondár, 2011). In turn, access to smaller earthquakes improves statistical seismology studies by providing more data and thus better insights into the underlying crustal processes. Interestingly, micro-seismicity has been shown to be critical for the observation of seismic precursors prior to large and potentially damaging earthquakes (Jones et al., 1982; Bernard et al., 1997; Wang et al., 2006; Ebel, 2008; Mignan, 2014; Brodsky, 2019; Trugman and Ross, 2019). Mignan (2014) proposed the following empirical law based on a meta-analysis of earthquake precursor studies:

$$m_{\min} < M - 3.0 \tag{1}$$

where $m_{\min}$ is the minimum magnitude threshold necessary to start observing seismic precursors prior to a mainshock of magnitude $M$. The completeness magnitude $M_c$ is the simplest, and one of the best proxies to the detection capability of a seismic network. Earthquake data, considered complete above this limit, follow the Gutenberg-Richter frequency-magnitude distribution. The optimal use of data requires that $m_{\min} = M_c$. As of 2011, the minimum $M_c$ in Mainland China was about 3.7 (Mignan et al., 2013).

In the past ten years, the detection capability, or $M_c$, of the seismic network in Mainland China has been improved with the upgrading of processing techniques (e.g., Wang et al., 2017), while the basic network layout remains unchanged. The China Earthquake Administration (CEA) has recently launched an ambitious nationwide network construction project that aims to extend the scale and applications of the seismic network with a total investment of approximately 1.87 billion RMB (290 million USD; https://www.cenc.ac.cn/cenc/zt/361404/361414/361563/index.html, last accessed: January 2022). Through this project, the number of stations in the seismic network will increase from approximately 2,000 to 15,000 in the future (Figure S1), with the data planned to be transmitted in real-time. The customary assessment methods (e.g., Mignan and Woessner, 2012) to estimate the $M_c$ are hampered for this network at the planning, design and implementation stages by a lack of observations.

The Bayesian Magnitude of Completeness (BMC) method proposed by Mignan et al. (2011) provides a promising solution to address the above problem of lack of observations to estimate $M_c$ by using Bayesian inference relating the spatial density of stations in the network to $M_c$ values, weighted by the respective uncertainties of the observations and the prior model. This method has been successfully applied in observed induced (Mignan, 2021) and natural seismicity, e.g., Taiwan (Mignan et al., 2011), Mainland China (Mignan et al., 2013), Switzerland (Kraft et al., 2013), Lesser Antilles arc (Vorobieva et al., 2013), California (Tormann et al., 2014), Greece (Mignan and Chouliaras, 2014), Iceland (Panzera et al., 2017), South Africa (Brandt, 2019) and Venezuela (https://www.statistics.gov.hk/wsc/CPS204-P47-S.pdf, last accessed: September 2021).

As an updating of the analysis applying BMC method to Mainland China by Mignan et al. (2013), we begin this article with a evaluation of detectability capability based on the earthquake catalog accumulated from 2012 to 2021 inside Mainland China. Once the BMC method is calibrated for the existing seismic network, we then extend its application to predict the future performance of the planned network. Finally, we quantify the potential improvement of seismicity in a more complete catalog observed with the change in the network, and investigate the significance for seismic-based earthquake prediction research based on the optimal use of data, namely $m_{\min} = M_c$. Our work will serve as an important reference to guide the design and optimization of the planned seismic network upgrade in Mainland China.

## 2 Data

2.1 Earthquake catalog (2012-2021)

For our purpose, we define the inside of Mainland China as the region within 100 km outside of the boundaries of Mainland China and Hainan Island (approximately 11.5 million km$^2$), where the China Earthquake Networks Center (CENC) produces a catalog based on the same standard and workflow. Approximately 890,000 earthquakes, in which 18,000 with $M_L \geq$ 3.0, that occurred inside Mainland China were reported in the catalog provided by the CENC from January 1, 2012, to July 11, 2021. Among them, the largest and the most damaging earthquake was the 2008 $M_S$ 8.0 Wenchuan earthquake, which caused approximately 70,000 deaths and 20,000 missing people.

2.2 The seismic networks

The existing seismic network in Mainland China consists of a high-gain broadband seismic network and a strong-motion network with a total of approximately 950 and 1,100 stations, respectively (Figure S1a). The new seismic network is being constructed by the National System for Fast Report of Intensities and Earthquake Early Warning project of China. This project is led by the CEA and was launched and implemented in 2015 and 2018, respectively. The new network consists of a total of approximately 15,000 stations: approximately 2,000 datum stations equipped with three-component broadband seismometers and accelerometers, approximately 3,200 basic stations equipped with only three-component accelerometers, and approximately 10,000 ordinary stations equipped with low-cost micro-electro-mechanical system (MEMS) intensity sensors (Figure S1b). It is expected that, by end of 2022, the new network will officially be online (Wenhui Huang, Qiang Ma and Changsheng Jiang, written comm., 2021). Only the data recorded by the broadband seismic stations in the existing network and the datum stations in the planned network are used to compile catalogs.

We computed the inter-station distance from the average distance of a given site to its closest fourth broadband and/or strong-motion stations (Kuyuk and Allen, 2013; Li et al., 2016; Li et al., 2021) using a mesh of 0.5°×0.5° resolution. In 40% of Mainland China, the inter-station distances of the existing network are less than 100 km, 7.8% are 50 km, and 0.4% are 25 km (Figures S2a and S2c). In contrast, in half of Mainland China, the inter-station distance will soon be smaller than 100 km, 23% be 50 km, and 3% be 25 km (Figure S2b and S2c). Although the area with an inter-station distance smaller than 100 km to be covered by the new network is almost the same as that covered by the old network, the areas with an inter-station distance smaller than 50 km and 25 km are expanded by factors of 2 and 7, respectively. The density of the new network is obviously improved. The new network will have an inter-station distance of approximately 20-30 km in North China, the central China north-south seismic belt, the southeast coast, the middle section of the Tianshan Mountains in Xinjiang, and approximately 10 km around large urban agglomerations, e.g., the center of Beijing Capital Circle, southeast coastal city cluster, Lanzhou-Xi'an, Urumqi, Chengdu, and Kunming (Figure S2b). Overall, the density of the new network in North China, the central China north-south seismic belt, the southeast coast, the middle section of the Tianshan Mountains in Xinjiang will be comparable to the most density seismic network worldwide, e.g., Japan, the U.S. west coast, and Taiwan (e.g., Kuyuk and Allen, 2013).

# 3 Monitoring capability estimated by $M_c$

3.1 Past network performance

Figure 1a shows $M_c$ maps for time periods from 2012 to 2021. We use the standard frequency-magnitude distribution-based mapping method (Wiemer and Wyss, 2000) with grid resolution of 0.5°×0.5° for each of them, and define a cylinder with a fixed radius of $r = 50$ km and a minimum number of earthquakes $N_{min} = 50$. The bulk $M_c$ in each grid is computed based on the events within a cylinder by using the non-parametric median-based analysis of the segment slope technique (MBASS; Amorèse, 2007), and its deviation is obtained from 200 bootstrap samples (Efron, 1979). Many gaps remain in grid points with low seismicity within the cylinder. Simply increasing $r$ is not recommended due to possible over-smoothing with obvious artifact patterns (Mignan and Woessner, 2012). The BMC technique is based on the density of the seismic network for determining $r$ (Mignan et al., 2011). This method efficiently decreases the number of gaps, but it can be used only where network information is available.

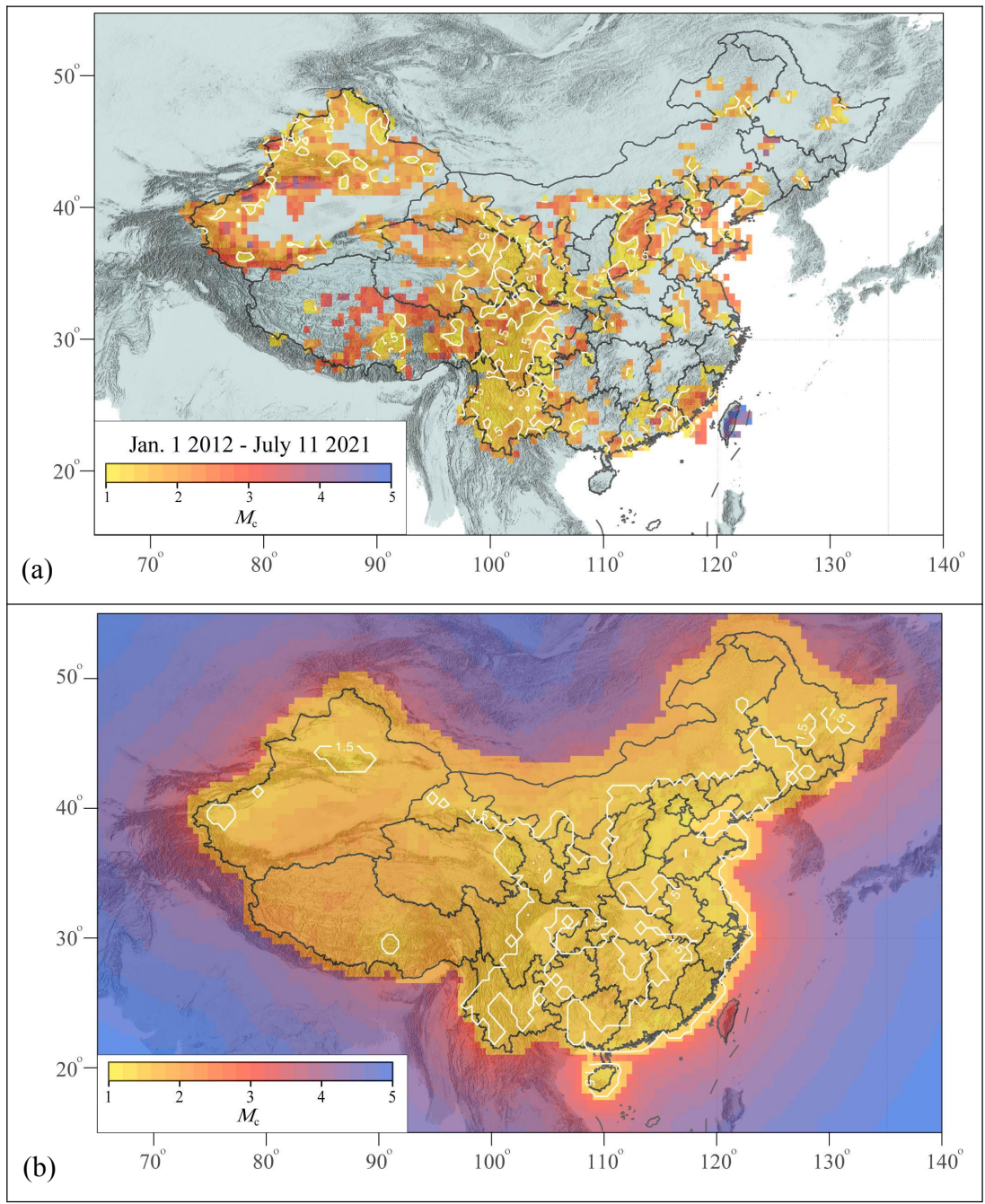

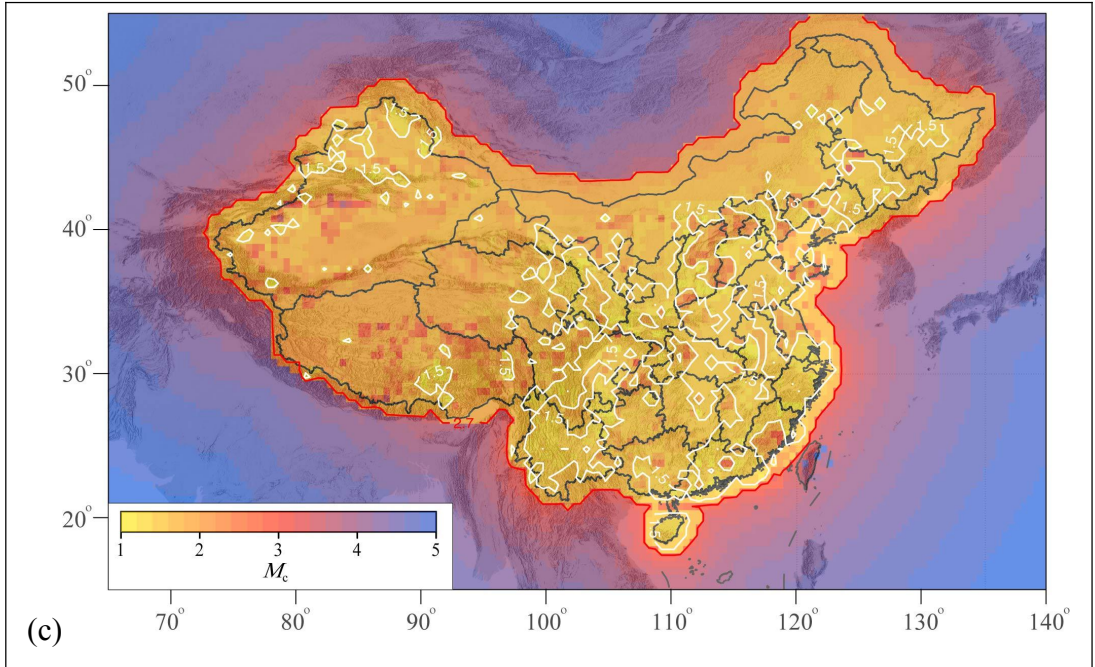

**Figure 1.** (a) Observed $M_c$ maps for time periods from 2012 to 2021. The median $M_c$ based on the $M_c$ spatial distribution is 1.8. The $M_c$ estimates are computed by using the median-based analysis of the segment slope (MBASS; Amorèse, 2007) and based on the standard frequency-magnitude distribution-based mapping method (Wiemer and Wyss, 2000). (b) Predicted $M_c$ map based on prior information for the existing broadband seismic stations in Figure S1a. The standard deviations $\sigma$ inside and outside Mainland China are 0.47 and 0.24, respectively. (c) Posterior $M_c$ map estimated by the Bayesian magnitude of completeness (BMC) method for time periods from 2012 to 2021. White and red lines show the $M_c = 1.5$ and $M_c^{post} = 2.7$ contours, respectively.

The overall spatial $M_c$ inside Mainland China in Figure 1a is basically the same as that from October 1, 2008, to 31 August 2011 of Mignan et al. (2013), which estimated a median $M_c$ value of 1.6. The median $M_c$ values based on its spatial distribution in Figure 1a is 1.8. Figure 1a produced by the standard mapping method also exhibits many gaps in regions with low seismicity where $M_c$ cannot be computed with confidence. The $M_c$ values outside Mainland China (e.g., Taiwan) are larger, because in this case, earthquakes were located by the CENC using only the national broadband seismic network. The areas with $M_c$ values smaller than 1.5 account for 17% of the total inside Mainland China. Areas with $M_c$ values smaller than 1.5 are sparse in the central China north-south seismic belt, northern part of Xinjiang, North China, Southeast China, and Lhasa.

3.2 Bayesian Magnitude of Completeness (BMC) application

The BMC method proposed by Mignan et al. (2011) merges the observed $M_c$ with prior Bayesian information, which is deduced from the relationship between the density of the seismic network and $M_c$ observations, weighted by their respective standard deviations. The posterior $M_c$ is obtained in a two-step procedure. First, a spatial resolution optimization is implemented. However, we use the standard mapping method with fixed parameters to obtain the spatial distribution of $M_c$ to avoid the iterated use of the network information in observed $M_c$ and prior Bayesian information, at the cost of the non-optimization of the spatial resolution (in contrast to previous BMC applications in other regions). Therefore, only the standard frequency-magnitude distribution-based mapping method is adopted in this article to produce Figure 1a. In addition, we still use the MBASS technique instead of the maximum curvature (MAXC) approach to consider the possible $M_c$ heterogeneities introduced by the aforementioned choice. The MAXC approach is a standard method of estimating the observed $M_c$ in the BMC method while resulting in large uncertainty, e.g., $\sigma = 0.56$ in Mignan et al. (2013) for the Chinese catalog. The MBASS technique is here shown to reduce those uncertainties in the context of this study. Second, a Bayesian approach to merge the observed $M_c^{obs}$ and the prior $M_c^{pred}$ is completed using Gaussian conjugates, giving the posterior $M_c^{post}$ for each grid as

$$M_c^{post} = \frac{M_c^{pred}\sigma_0^2 + M_c^{obs}\sigma^2}{\sigma_0^2 + \sigma^2} \quad (2)$$

in which $\sigma$ and $\sigma_0$ are the standard deviations for $M_c^{pred}$ and $M_c^{obs}$, respectively. The posterior standard deviation $\sigma_{post}$ can be estimated as

$$\sigma_{post} = \sqrt{\frac{\sigma_0^2 \sigma^2}{\sigma_0^2 + \sigma^2}} \quad (3)$$

For the prior model, $M_c^{pred} = f[d(4)]$, in which $d(4)$ is the distance to the fourth closest station that is used to fit the observed $M_c$. As a novelty, we compare three models: (1) $M_c^{pred} = a \cdot d(4) + b$, with a degree-of-freedom (df) of 2, (2) $M_c^{pred} = a \cdot d(4)^c + b$, with a df of 3, and (3) $M_c^{pred} = a \cdot \log_{10} d(4) + b$, with a df of 2. The observed grid value $M_c$ in every year from 2012 to 2021 is used to evaluate the alternative models. The four indicators are the sum of squares due to error (SSE), the coefficient of determination (R-square), the root mean squared error (RMSE) and the Akaike information criterion (AIC; Akaike, 1970). We found that the model $M_c^{pred} = a \cdot d(4)^c + b$ performs best in terms of both the goodness of fit (SSE = 1438, R-square = 0.0755, and RMSE = 0.4695) and model evaluation (AIC = -9862.8) (Table S1). This model is exactly the one used in the former BMC, e.g., for Taiwan (Mignan et al., 2011). Figure 1b shows the predicted $M_c$ map based on a prior model for the existing broadband seismic stations. The standard deviations $\sigma$ inside Mainland China are 0.47.

Figure 1c shows the spatial distribution of the posterior $M_c$ estimated by the BMC technique. The distribution of the posterior $M_c$ is more heterogeneous than that of the prior $M_c$ since the latter is a continuous function of the density of stations. The minimum posterior $M_c$ value in Mainland China was about 2.7. The areas with posterior $M_c$ values smaller than 1.5 and 2.0 account for 38% and 91% inside Mainland China, respectively, and are 8% lower for both of them than for the prior $M_c$ values, namely, 1.5 for 46% and 2.0 for 99% (Figure 2a). The area differences larger than 5% at $M_c$ = 1.4 and 1.5 (Figure 2a) can be explained by the relatively large uncertainty ($\sigma$ = 0.47) of the prior model compared with that for Switzerland ($\sigma$ = 0.16), Taiwan ($\sigma$ = 0.18), South Africa ($\sigma$ = 0.18), Venezuela ($\sigma$ = 0.19), Greece ($\sigma$ = 0.29), Iceland ($\sigma$ = 0.3) and induced seismicity ($\sigma$ = 0.37) (see matching references in the introduction). However, our $\sigma$ is smaller than that computed by Mignan et al. (2013) for the Chinese catalog from 2008 to 2011, in which $\sigma$ = 0.56.

### 3.3 Bayesian Magnitude of Completeness (BMC) application

With the calibration of the BMC prior from the observed $M_c$, we can then predict the $M_c$ distribution for any seismic network configuration, assuming that the prior model is valid when used in the same area (i.e., same seismic waveform attenuation laws). Figure 2b shows the predicted $M_c$ distribution for the planned broadband seismic network. Figure 2a shows the increase in the spatial distribution of the regions with $M_c$ values smaller than 1.5. Quantitatively compared with the posterior $M_c$ of the existing network (Figure 1c), the area with $M_c \leq 1.5$ expands by a factor of 1.5 to reach 54%, and the $M_c$'s of almost the entire inside Mainland China (99%) are less than 2.0. The monitoring capability in Xinjiang increases dramatically, followed by the central China north-south seismic belt and North China. Southeast China shows the least gain since the existing configuration is currently sufficient with only a small improvement of monitoring capability needed.

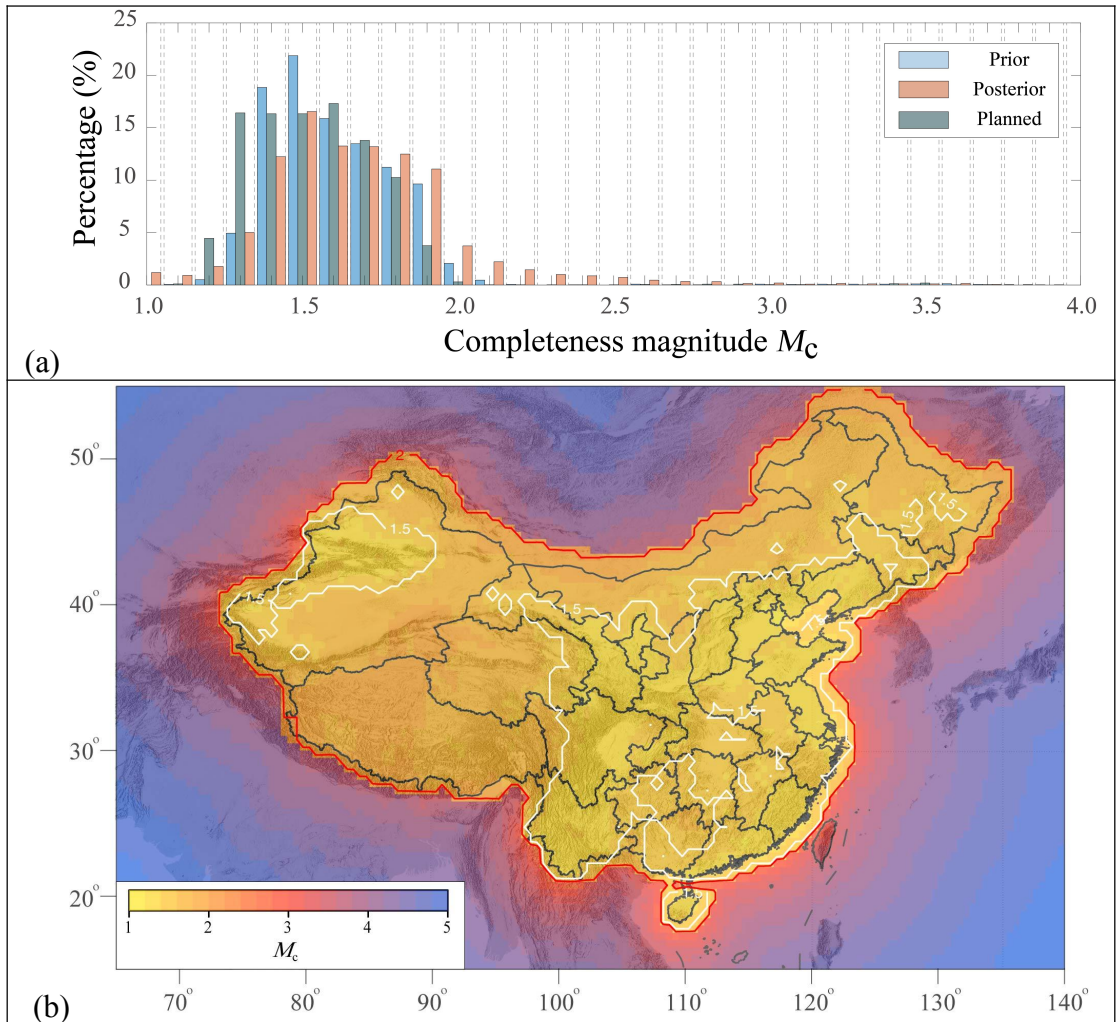

**Figure 2.** (a) The binned distributions in prior (Figure 1b) and posterior (Figure 1c), and predicted $M_c$ values inside Mainland China. (b) Predicted $M_c$ map based on prior information for the planned datum stations. The standard deviations $\sigma$ inside and outside Mainland China are same as Figure 1b. White and red lines show the $M_c = 1.5$ and $M_c^{post} = 2.0$ contours, respectively.

## 4 Benefits from a more complete catalog

Better monitoring capability leads to the detection of more small earthquakes by offering a higher resolution for active fault network structures (e.g., Ouillon and Sornette, 2011), clarifying the relevant components of stress that would trigger subsequent earthquakes (e.g., Mancini et al., 2019), and helping to understand the earthquakes' clustering behavior (e.g., Sornette and Werner, 2005a; 2005b) and the physics of the Earth's crust (e.g., Mignan, 2011). Indeed, according to current understanding, as a whole, small earthquakes contribute the most in the triggering of other earthquakes, including medium size to large earthquakes (Helmstetter, 2003; Helmstetter et al., 2005; Nandan et al., 2021). It is therefore of the upmost importance to record them, as much as possible (Ebel, 2008; Brodsky, 2019; Trugman and Ross, 2019).

The 99th percentiles of the posterior $M_c$ for the existing network (Figure 1c) and prediction (prior) $M_c$ for the planned network (Figure 2b) are 2.7 and 2.0, respectively. Statistically, 29,413 earthquakes with $M \geq 2.7$ were observed (on average, approximately 3,000 per year and 250 per month) from January 2012 to July 2021. According to the Gutenberg-Richter law that fits the empirical magnitude distribution for magnitudes larger than 2.7 ($a$-value of 6.76 and $b$-value of 0.85), approximately 500 earthquakes larger than 2.0 that are missed by the existing network would be recorded in the planned network. These values suggest that the available seismicity samples (approximately 115,800) will increase by a factor of at least 3. We predict that the complete catalog of the planned broadband seismic network will record approximately 12,100 and 1,000 earthquakes per year and month, respectively, which are both at least a three-fold increase compared with the number of earthquakes recorded by the existing network. The improvement would potentially strengthen both predictive skills of the statistical and physical-based models because the sample number of earthquakes would be increased.

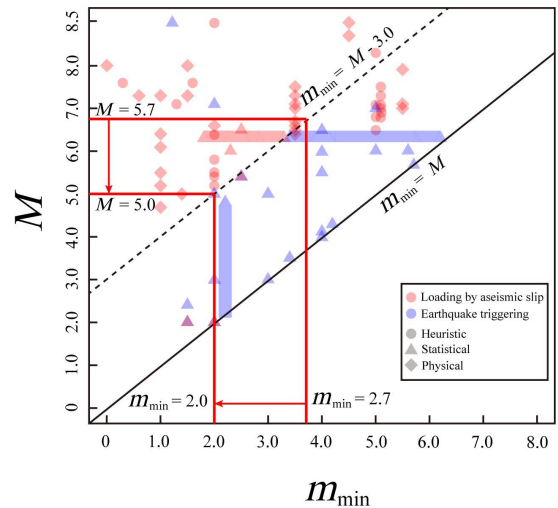

**Figure 3.** Demonstration of the decrease of mainshock magnitude $M$ from minimum foreshock magnitude $m_{min} = 2.7$ for the existing network to $m_{min} = 2.0$ for the new network inside Mainland China. A total of 77 data points as the background are from the meta-analysis of 37 published studies by Mignan (2014), in which arguments are mainly based on heuristic, statistical or physical considerations. Modified from Figure 2 of Mignan (2014).

Equation (1) that derives from a meta-analysis of 37 foreshock studies published from 1982 to 2013 shows that seismic precursors are most likely to be observed at least at 3 units below the mainshock magnitude $M$ (Figure 3). Figures 1c and 2b show that the minimum $M_c$ inside Mainland China is 2.7 for the existing network and 2.0 for the new network, respectively. Considering the optimal use of data, namely $m_{min} = M_c$, we find that the magnitude $M$ of mainshocks prior to which anomalous foreshock activity most likely to be found to emerge is decreased from 5.7 for the existing network to 5.0 for the future network. Assuming that earthquakes in Mainland China are potentially destructive for magnitude $M$ larger than 5, the future seismic network will achieve the goal of almost full coverage for optimal seismic precursor-based earthquake prediction research.

## 5 Conclusions and discussion

The performance of the network will significantly improve by simply densifying the existing layout. In half of Mainland China, the inter-station distance will soon be smaller than 100 km (and smaller than 50 km and 25 km for 23% and 3% of the mainland, respectively). The area with an inter-station distance smaller than 50 km and 25 km that are to be covered by the new network are expanded by factors 2 and 7, respectively. In this study, we quantified the higher-resolution detection of the frequent smaller earthquakes inside Mainland China via the completeness magnitude ($M_c$) metric for the existing and planned seismic networks. Using the best performance prior model ($\sigma = 0.47$) of the Bayesian Magnitude of Completeness (BMC) method calibrated on the Chinese catalog from January 1, 2012 to July 11, 2021, we predicted the spatial distribution of $M_c$ for the future network based on its network configuration. Compared with the posterior $M_c$ of the existing network, the area with $M_c \leq 1.5$ that is covered by the new network expands by a factor of 1.5 to reach 54%. If 95% of Mainland China is at present covered down to $M_c = 2.7$, this value will soon fall to $M_c = 2.0$. We predicted that the complete catalog of the planned network will record a factor of three more earthquakes per year compared with that of the existing network (for $a = 6.76$ and $b = 0.85$ in the Gutenberg-Richter law $\log_{10}N = a - b \cdot M_c$). Taking the minimum $M_c$ inside Mainland China as 2.0 for the future network and assuming earthquakes to be potentially damaging at $M \geq 5$, the new network shall achieve the goal of almost total coverage for optimal seismic-based earthquake prediction research in the entire Mainland China. Our investigation provides a useful reference for the real functioning and further optimization of the new networks in Mainland China

The improvement in computing power makes the adoption of artificial intelligence (AI) in the routine monitoring of seismic networks possible. Previous tests showed that more small earthquakes by at least a factor of ten could be detected by AI compared with that produced by the current processing techniques (e.g., Beroza et al., 2021). Therefore, we can expect that the monitoring capabilities of the seismic network will usher a comprehensive and large improvement in the near future with the revolution in seismic observational technology enabled by AI and the further upgradation of the software in the near future.

**Acknowledgments**

The authors would like to thank Dr. Yang Zang, Dr. Zhengyi Yuan, Dr. Chen Yang, and Dr. Lei Tian at the China Earthquake Networks Center; Dr. Ke Sun, and Dr. Zijian Cui at the Institute of Earthquake Forecasting of China Earthquake Administration for providing the earthquake catalog and information of stations. This project is supported by the National Natural Science Foundation of China under Grant No. U2039202 and the Guangdong Basic and Applied Basic Research Foundation under Grant No. 2020A1515110844.

**Data Availability Statement**

The earthquake catalog and information of the existing and the planned seismic networks were provided by China Earthquake Networks Center (doi: 10.11998/SeisDmc/SN) through the internal link (Earthquake Cataloging System at CEA: http://10.5.160.18/console/index.action). The Bayesian magnitude of completeness method used in the study is from Mignan et al., (2011).

Supporting Information for

Predicting the Future Performance of the Planned Seismic Network in Mainland China

Jiawei Li[1], Arnaud Mignan[1,2], Didier Sornette[1,3], and Yu Feng[1]

[1]Institute of Risk Analysis, Prediction and Management (Risks-X), Academy for Advanced Interdisciplinary Studies, Southern University of Science and Technology (SUSTech), Shenzhen, China.

[2]Department of Earth and Space Sciences, Southern University of Science and Technology (SUSTech), Shenzhen, China.

[3]Department of Management, Technology and Economics (D-MTEC), Swiss Federal Institute of Technology in Zürich (ETH Zürich), Zürich, Switzerland

**Contents of this file**

Figures S1 to Existing (as of 2018) and planned (as of 2022) stations in Mainland China.

Figures S1 to Maps showing the inter–station distances for the existing and planned networks in China.

Tables S1 to Characterization parameters of alternative models and the results of their evaluation to observed $M_c$ values.

## Introduction

Using the Bayesian Magnitude of Completeness (BMC) method, we generated the spatial distribution of Mc predicted for the new network, based on the prior model calibrated on the current earthquake catalog from January 1, 2012, to July 11, 2021, and network configuration. We further quantify the potential improvement of seismicity in a more complete catalog observed with the change in the network and investigate the significance for seismic-based earthquake prediction research based on the optimal use of data. As an important supplement to the text, this file contains plots of existing and planned stations in Mainland China and maps showing the inter-station distances for the existing and planned networks. Characterization parameters of alternative models and the results of their evaluation to observed Mc values are also shown here.

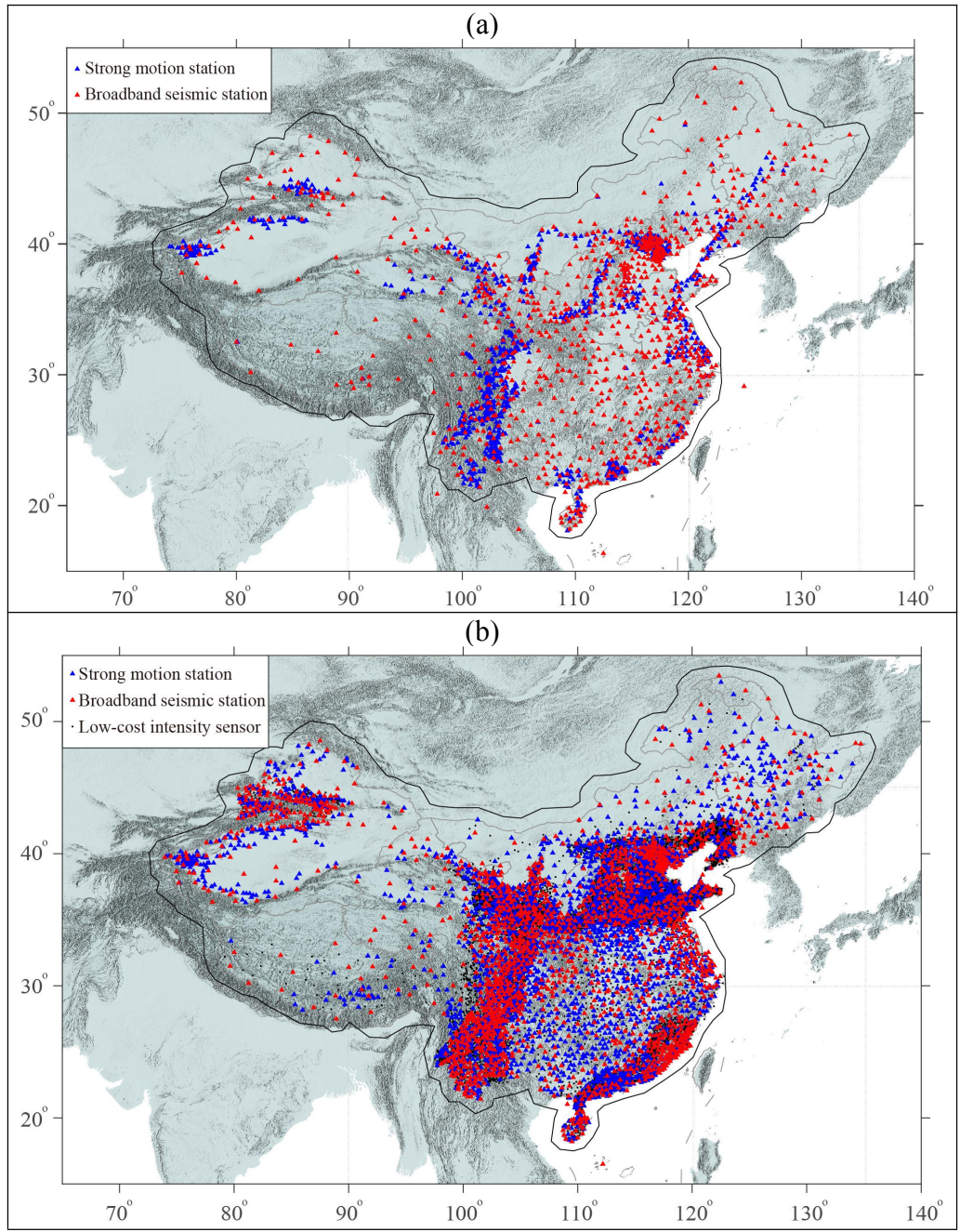

**Figure S1.** (a) Existing (as of 2018) stations as part of the China Earthquake Networks Center (CENC) and the China Strong Motion Network Center (CSMNC). (b) Planned (as of 2022) stations to be newly deployed or upgraded in the National System for Fast Report of Intensities and Earthquake Early Warning of China. We define 'inside' in this study (black line) as the region within 100 km outside of Mainland China's boundaries.

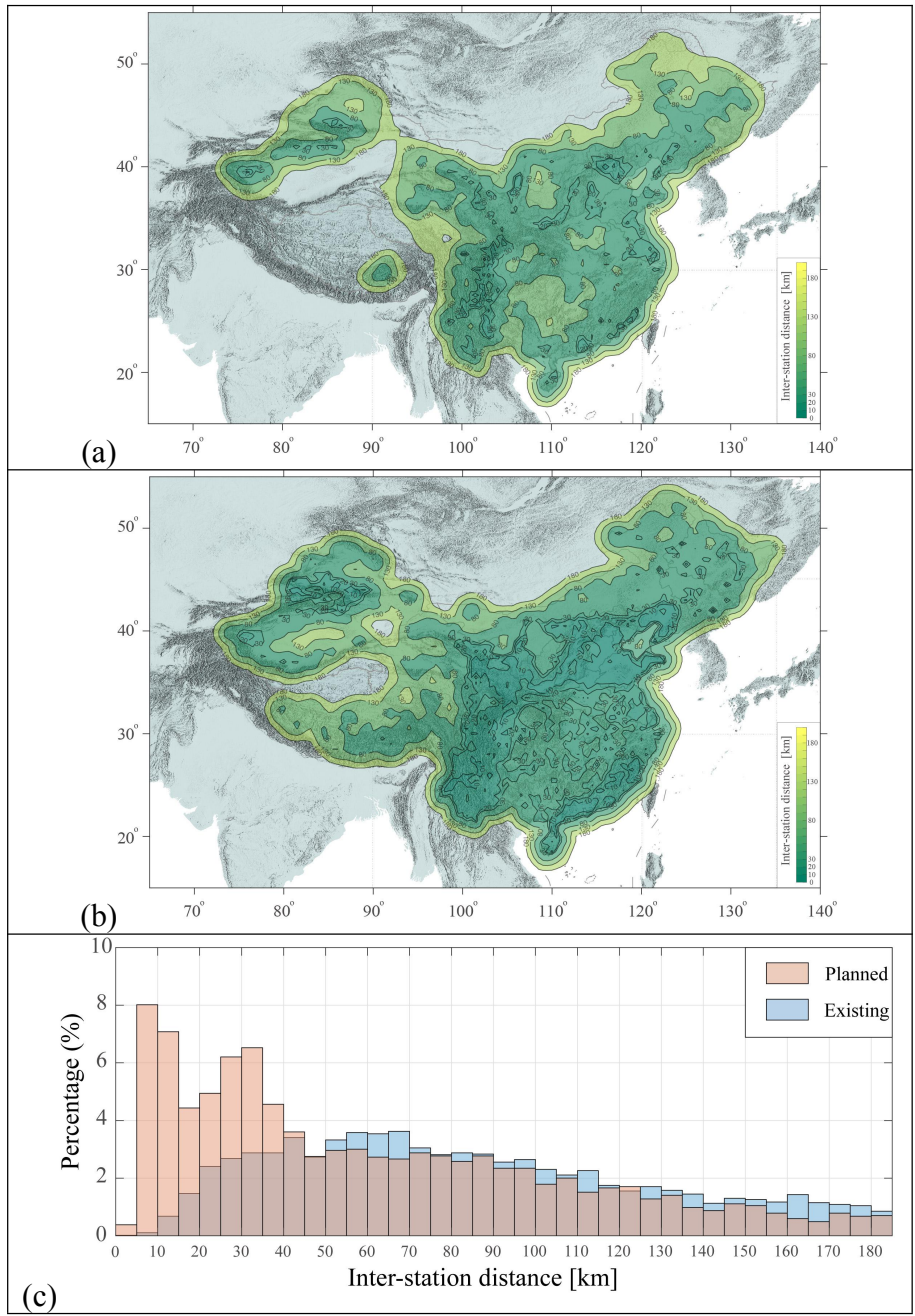

**Figure S2.** Maps showing the inter-station distances for the (a) existing and (b) planned networks. For any given site (assuming a grid of 0.5° × 0.5° resolution). The inter-station distance is calculated as the average distance to the four closest stations (Kuyuk and Allen, 2013; Li et al., 2016; Li et al., 2021). The histogram in (c) shows the binned distribution of the inter-station distances inside Mainland China.

Table S1. Characterization parameters of alternative models and the results of their evaluation to observed $M_c$ values.

| Model | a | b | c | df | SSE | R-square | RMSE | AIC |
|---|---|---|---|---|---|---|---|---|
| $y = a \cdot d(4) + b$ | 0.002 | 1.255 | -- | 2 | 1448 | 0.0687 | 0.4712 | −9819.6 |
| $y = a \cdot d(4)^c + b$ | 0.128 | 0.767 | 0.365 | 3 | 1438 | 0.0755 | 0.4695 | −9862.8 |
| $y = a \cdot \log_{10} d(4) + b$ | 0.602 | 0.261 | -- | 2 | 1442 | 0.0731 | 0.4701 | −9846.7 |

$d(4)$: distance to the closest fourth station; df: degree-of-freedom; SSE: sum of squares due to error; R-square: coefficient of determination; RMSE: root mean square error; AIC: Akaike information criterion.